\documentclass[aps,preprint,nofootinbib,superscriptaddress]{revtex4}
\usepackage{amssymb}

\usepackage{epsfig}
\usepackage[bookmarksnumbered,bookmarksopen,colorlinks,citecolor=blue,linkcolor=blue]{hyperref}
\begin{document}


\title{Modification of mass formula by considering isospin effects}

\author{Ning Wang}
\email{wangning@gxnu.edu.cn}  \affiliation{Department of Physics,
Guangxi Normal University, Guilin 541004, P. R. China}

\author{Min Liu}
\affiliation{ College of Nuclear Science and Technology, Beijing
Normal University, Beijing, 100875, P. R. China}
\affiliation{Department of Physics, Guangxi Normal University,
Guilin 541004, P. R. China}

\author{Xizhen Wu}
\affiliation{China Institute of Atomic Energy, Beijing 102413, P.
R. China}

\begin{abstract}
We propose a semi-empirical nuclear mass formula based on the
macroscopic-microscopic method in which the isospin and mass
dependence of model parameters are investigated with the Skyrme
energy density functional. The number of model parameters is
considerably reduced compared with the finite range droplet model.
The rms deviation with respect to 2149 measured nuclear masses is
reduced by 21\%, falls to 0.516 MeV. The new magic number $N=16$
in light neutron-rich nuclei and the shape coexistence phenomena
for some nuclei have been examined with the model. The shell
corrections of super-heavy nuclei are also predicted.
\end{abstract}

\maketitle

\begin{center}
\textbf{I. INTRODUCTION}
\end{center}

The nuclear mass is of great importance not only for various
aspects of nuclear physics, but also for weak-interaction studies
and astrophysics \cite{Lun03}. In nuclear physics, it is helpful
to study the nuclear symmetry energy and the synthesis of
super-heavy nuclei by considering the more than 2000 measured
nuclear masses. Theoretically, the mass of an atomic nucleus can
be calculated by the macroscopic-microscopic method (such as the
finite-range droplet model \cite{Moll95})  or the microscopic
approaches (such as the Hartree-Fock Bogoliubov approach
\cite{HFB14,HFB17}) or some other mass formulas \cite{DZ}. The
best mass formulas at present can reach about 0.6 MeV in the rms
deviation for the usual data set of 2149 measured masses of nuclei
(N and $Z \ge 8$) \cite{Audi} with about $24\sim30$ model
parameters. Compared with the microscopic Hartree-Fock (HF)
approaches, the macro-micro model is much faster in the
calculation of the nuclear masses for the whole nuclear chart
which provides a possibility for performing a large scale nuclear
mass calculations to refine the model parameters and to explore
the global behavior of nuclei. However, there are two crucial
points in the macro-micro method should be further studied. The
first one is that the consistency of the model parameters between
the macroscopic and microscopic parts in the macro-micro method
should be improved. It is known that although the finite-range
droplet model (FRDM) is widely used in the calculations of nuclear
mass, the parameter values in the calculation of the microscopic
shell corrections are different from the corresponding values used
in the macroscopic part of the model \cite{Lun03}. This less
consistency between the macroscopic and microscopic parts may
considerably reduce the credibility of extrapolations of the
macroscopic-microscopic approach. On the other hand, with the
great development of the experimental facilities for the study on
super-heavy nuclei and nuclei far from the $\beta$-stability line,
the influence of isospin effects on the nuclear mass formula
attracted a great attention and should be given a better
consideration. Based on above discussions an improved nuclear mass
formula of self-consistently considering the isospin effects in
both macroscopic and microscopic parts would be necessary to be
established for providing a large scale nuclear mass calculations.

To investigate the consistency of the model parameters between the
macroscopic and microscopic parts in the macroscopic-microscopic
approach and isospin dependence of the model parameters, the
Skyrme energy density functional approach together with the
extended Thomas-Fermi (ETF) approximation \cite{Liu06,Bart02} is
used. It is known that the energy density functional theory is
widely used in the study of the nuclear ground state which
provides us with a useful balance between accuracy and computation
cost allowing large systems with a simple self-consistent manner.
With the Skyrme energy density functional approach, we
systematically investigate some ground state properties of nuclei,
such as the nuclear symmetry energy coefficient, the deformation
energy and the symmetry potential, which are helpful to improve
the macro-micro method. Based on these calculations, we propose a
semi-empirical nuclear mass formula by taking into account the
isospin and mass dependent model parameters. The paper is
organized as follows: In Sec. II, the proposed mass formula is
introduced. In Sec. III, some calculation results are presented.
Finally, a summary is given in Sec. IV.

\begin{center}
\textbf{II. THE MODEL}
\end{center}

In this section, we first introduce the macroscopic part of the
mass formula. Then, the influence of nuclear deformation on the
macroscopic energy of nucleus are investigated with the Skyrme
energy density functional approach and the single particle
potential used in the calculation of the microscopic shell
correction is introduced. In addition, the symmetry potential and
the symmetry energy coefficient of nuclear matter is also
investigated. Finally, the parameters adopted in the model are
presented.

\begin{center}
\textbf{A. Modified Bethe-Weizs\"acker Mass Formula}
\end{center}

We start with the macroscopic-microscopic method
\cite{Moll95,Cohen}. The total energy of a nucleus can be calculated
as a sum of the liquid-drop energy and the Strutinsky shell
correction $\Delta E$,
\begin{eqnarray}
E (A,Z,\beta)=E_{\rm LD}(A,Z) \prod_{k \ge 2} \left (1+b_k \beta_k^2
\right )+\Delta E (A,Z,\beta).
\end{eqnarray}
The liquid drop energy of a spherical nucleus $E_{\rm LD}(A,Z)$ is
described by a modified Bethe-Weizs\"acker mass formula
\cite{Heyde},
\begin{eqnarray}
E_{\rm LD}(A,Z)=a_{v} A + a_{s} A^{2/3}+ a_{c}
\frac{Z(Z-1)}{A^{1/3}} \left ( 1- Z^{-2/3} \right) + a_{sym} I^2 A
+ a_{pair}  A^{-1/3}\delta_{np}
\end{eqnarray}
with isospin asymmetry $I=(N-Z)/A$. The pairing term proposed in
\cite{Tem} is adopted, with
\begin{eqnarray}
\delta_{np}= \left\{
\begin{array} {r@{\quad:\quad}l}
  2 - |I|  &   N {\rm ~and~} Z {\rm ~even }    \\
      |I|  &   N {\rm ~and~} Z {\rm ~ odd }    \\
  1 - |I|  &   N {\rm ~even,~} Z {\rm ~odd,~ } {\rm ~and~ } N>Z   \\
  1 - |I|  &   N {\rm ~odd,~} Z {\rm ~even,~ } {\rm ~and~ } N<Z   \\
  1             &   N {\rm ~even,~} Z {\rm ~odd,~ } {\rm ~and~ } N<Z   \\
  1             &   N {\rm ~odd,~} Z {\rm ~even,~ } {\rm ~and~ } N>Z   \\
\end{array} \right.
\end{eqnarray}

 In this work, the symmetry energy coefficient of finite nuclei
is written as,
\begin{eqnarray}
 a_{sym}=c_{sym}\left [1-\frac{\kappa}{A^{1/3}}+\frac{2-|I|}{ 2+|I|A} \ \right
 ],
\end{eqnarray}
based on the conventional surface-symmetry term
\cite{Pear01,Kirson} of liquid drop model, with a small correction
term for description of isospin dependence of $a_{sym}$. The
sensitive dependence of symmetry energy coefficient on the
asymmetry of nucleus, especially that $a_{sym}$ increases with
increasing proton fraction of the system is also found in
\cite{Sam}. The introduced $I$ correction term approximately
describes the Wigner effect \cite{Moll95} of heavy nuclei. For a
heavy nucleus near the $\beta$-stability line ($A|I| \gg 2 \gg
|I|$), the introduced $I$ term in $a_{sym}$ roughly leads to a
correction $ \propto |I|$ (known as Wigner term) to the binding
energy of the nucleus. Compared with the case without the $I$ term
being taken into account, the rms deviation of nuclear masses
defined as $ \sigma^{2}=\frac{1}{m}\sum \left ( M_{\rm exp}^{(i)}
- M_{\rm th}^{(i)} \right )^2$ from the measured masses AME2003
\cite{Audi} for the 2149 nuclei ($N$ and $Z \ge 8$) is reduced by
about $6\%$. Furthermore, we find when the isospin dependence of
symmetry energy coefficient is taken into account, the obtained
optimal $c_{sym}$ increases about 3 MeV and up to 29 MeV which is
close to the symmetry energy coefficient of nuclear matter at
saturation density obtained from the Skyrme energy density
functional.

The Coulomb exchange correction and surface diffuseness correction
to the Coulomb energy is approximately taken into account as the
term $Z^{-2/3}$. In addition, the terms $b_k$ in Eq.(1) which are
obtained according to the Skyrme energy density functional (the
detailed discussion is in the next subsection) describe the
contribution of nuclear deformation to the macroscopic energy.

\newpage

\begin{center}
\textbf{B. Influence of Nuclear Deformation on the Macroscopic
Energy}
\end{center}

\begin{figure}
\includegraphics[angle=-0,width=1.0\textwidth]{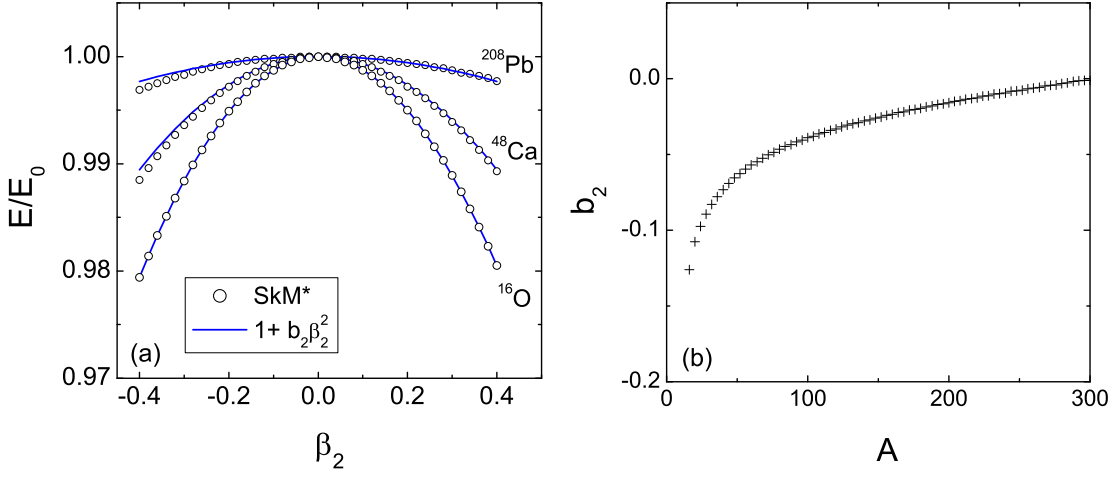}
 \caption{(Color online) (a) Energy of $^{16}$O, $^{48}$Ca and $^{208}$Pb with respect to $\beta_2$
deformation. Here, the values of $E_0$ are negative. The circles
and the solid curves denote the results of SkM* interaction and of
a formula $E/E_0=1+b_2\beta_2^2$, respectively. (b) The value of
$b_2$ obtained with SkM* as a function of mass number. }
\end{figure}
\begin{figure}
\includegraphics[angle=-0,width=1.0\textwidth]{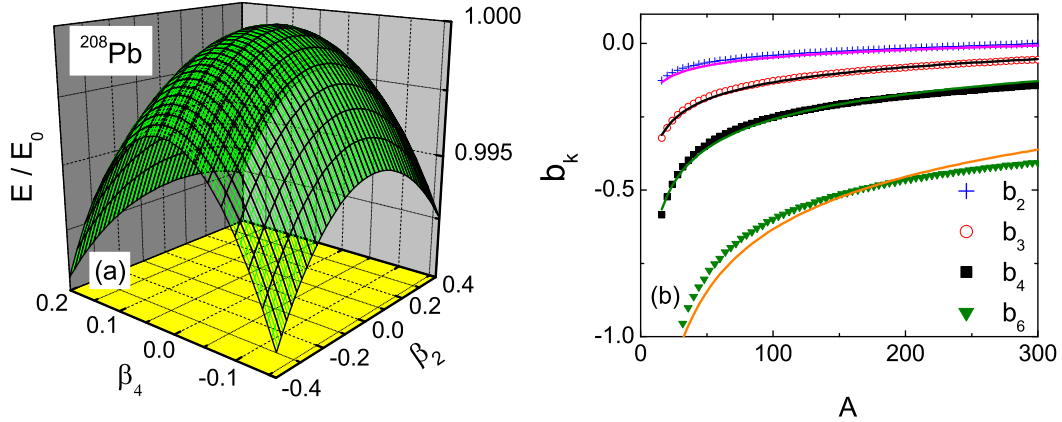}
 \caption{(Color online) (a) Energy of $^{208}$Pb with respect to $\beta_2$
 and $\beta_4$ deformation with Skyrme energy density functional approach.  (b) The value of $b_k$ as a function of mass number.
The scattered symbols denote the obtained curvatures of the
parabolas with the Skyrme force SkM* for a number of nuclei. The
solid curves denote the corresponding results with an empirical
formula (6).}
\end{figure}
For the deformation of nuclei, we only consider axially-deformed
cases. In this work, only $\beta_2$ and $\beta_4$ deformations of
nuclei are taken into account. We first investigated the energy of
a nucleus with respect to a $\beta_2$ deformation based on the
Skyrme energy density functional together with the extended
Thomas-Fermi approximation (ETF) \cite{Liu06,Bart02}. The
procedure is as follows: The total energy of a nucleus can be
expressed as the integral over the Skyrme energy density
functional $\mathcal H(\textbf{r})$ \cite{Bart82}. Given a density
functional $\rho(\textbf{r})$, one can calculate the corresponding
energy via $E = \int {\mathcal H[\rho(\textbf{r})]} d{\bf r}$
under the ETF approximation. We first obtain the binding energy
$E_0$ and the spherical Woods-Saxon density distributions of a
nucleus with the approach in \cite{Liu06}. Then, with the same
procedure we calculate the energy $E(\beta_2)$ of the nucleus with
a quadrupole deformed Woods-Saxon density distribution of the
nucleus in which the central density and the surface diffuseness
remained unchanged. Fig.1(a) shows the calculated energy of
$^{16}$O, $^{48}$Ca and $^{208}$Pb as a function of  $\beta_2$
deformation with the SkM* interaction \cite{Bart82} (denoted by
circles). The solid curves denote the results of a formula
$E/E_0=1+b_2 \beta_2^2$ in which the value of $b_2$ is obtained by
fitting the open circles. One can see that the parabola
approximation to the change of energy with $\beta_2$ is
acceptable. Fig.1(b) shows the value of $b_2$ as a function of the
mass number. The crosses denote the results of the SkM*
interaction for a number of nuclei along the $\beta$-stability
line. We find that the dependence of $b_2$ on the mass number $A$
 can be reasonably well described by a formula
\begin{eqnarray}
b_2=g_1A^{1/3}+g_2 A^{-1/3}.
\end{eqnarray}
This form of mass dependence of $b_2$ is therefore adopted in the
proposed mass formula and the optimal values of $g_1$ and $g_2$
are finally determined by the 2149 measured nuclear masses
\cite{Audi}.

To take into account the influence of the higher-multipole
deformation of nuclei, we investigate the change of energy of a
nucleus with respect to a certain set of nuclear deformation
parameters with the Skyrme energy density functional approach
mentioned previously. In Fig.2 (a), we show the energy of
$^{208}$Pb as a function of $\beta_2$ and $\beta_4$ with the
Skyrme force SkM*. We find that the influence of nuclear $\beta_4$
deformation on the nuclear energy can be roughly described by a
parabola at small deformations. For other higher-multipole
deformations, the parabola approximation can also be applied for
small deformation cases, and the proposed model can be easily
extended to consider other higher-multipole deformations.
Furthermore, we notice that the curvature of the parabola for a
given $\beta_k$ deformation can be approximately described by an
empirical formula
\begin{eqnarray}
b_k=\left ( \frac{k}{2} \right ) g_1A^{1/3}+\left ( \frac{k}{2}
\right )^2 g_2 A^{-1/3},
\end{eqnarray}
which is an extension of the formula (5). In Fig.2(b), we show the
mass dependent curvatures of the parabolas. The crosses, the open
circles, the solid squares and the triangles denote the obtained
curvatures $b_2$, $b_3$, $b_4$ and $b_6$ of the parabolas with the
Skyrme energy density functional approach for a number of nuclei
along the $\beta$-stability line, respectively. The solid curves
denote the corresponding results with the empirical formula (6)
taking $g_1=0.0074$ and $g_2=-0.38$. One can see that the
curvatures of the parabolas can be reasonably well described by
the empirical formula which greatly reduces the computation time
for the calculation of deformed nuclei.

\begin{center}
\textbf{C. Single-particle Potential in the Microscopic Part}
\end{center}

In the microscopic part, the shell correction
\begin{eqnarray}
\Delta E=c_1 E_{\rm sh},
\end{eqnarray}
is obtained by the traditional Strutinsky procedure \cite{Strut}
by setting the smoothing parameter $\gamma=1.2\hbar\omega_0$ and
the order $p=6$ of the Gauss-Hermite polynomials. Where, $E_{\rm
sh}=E_{\rm sh}(P)+E_{\rm sh}(N)$ i.e. sum of the shell energies of
protons and neutrons. $\hbar\omega_0=41 A^{-1/3}$MeV is the mean
distance between the gross-shells. In this work, we introduce a
scale factor $c_1$ to the shell correction.  This additional
parameter is used to adjust the division of the binding energy
between the macroscopic part and the remaining microscopic
correction. It is known that a similar scale factor is usually
introduced to the liquid-drop part \cite{Sag09} or the shell
correction part \cite{Shen08} to adjust the division between the
two parts for giving better results in the calculation of fission
barrier. It is necessary to investigate the influence of this
parameter on the nuclear masses. We find that the rms deviation
for the 2149 nuclear masses can be somewhat reduced with the
introduced factor $c_1$ compared with the case setting $c_1=1$.

To obtain the shell correction $\Delta E$, we execute a computer
code WSBETA \cite{Cwoik} to calculate the single particle levels
of an axially deformed Woods-Saxon potential and then perform the
Strutinsky procedure. The single-particle Hamiltonian in the code
WSBETA is written as
\begin{eqnarray}
H=T+V+V_{\rm s.o.},
\end{eqnarray}
with the spin-orbit potential
\begin{eqnarray}
V_{\rm s.o.}=-\lambda \left (\frac{\hbar}{2 M c} \right )^2\nabla
V \cdot (  \vec{ \sigma } \times  \vec{ p}),
\end{eqnarray}
where $\lambda $ denotes the strength of the spin-orbit potential.
In this work, we set $\lambda= \lambda_0 \left ( 1+\frac{N_i}{A}
\right )$ with $N_i=Z$ for protons and $N_i=N$ for neutrons. Here,
the isospin-dependent spin-orbit interaction strength is obtained
based on the Skyrme energy-density functional in which the
spin-orbit potential is usually expressed as
\begin{eqnarray}
V_{q}^{s.o.}= \frac{1}{2} W_0 \nabla (\rho+\rho_q) \cdot ( \vec{
\sigma } \times \vec{ p})\approx \frac{1}{2} W_0 \left (
1+\frac{N_i}{A} \right ) \nabla  \rho  \cdot ( \vec{ \sigma }
\times \vec{ p}),
\end{eqnarray}
with the nucleon density $\rho=\rho_p+\rho_n$  and the spin-orbit
strength $W_0$. $M$ in Eq.(10) is the free nucleonic mass, $\vec{
\sigma }$ and $\vec{ p}$ are the Pauli spin matrix and the nucleon
momentum, respectively \cite{Cwoik}. The central potential $V$ is
described by an axially deformed Woods-Saxon form
\begin{eqnarray}
V (\vec {  r} \, )= \frac{V_q}{1+  \exp (\frac{  r -\mathcal {R}
(\theta)}{a})}.
\end{eqnarray}
Where,  the depth $V_q$ of the central potential ($q=p$ for
protons and $q=n$ for neutrons) is written as
\begin{eqnarray}
 V_q = V_0 \pm V_s I
\end{eqnarray}
with the plus sign for neutrons and the minus sign for protons.
$V_s$ is the isospin-asymmetric part of the potential depth. We
assume $V_s=a_{sym}$ in this work (detailed study of the relation
between $V_s$ and $a_{sym}$ is given in the following part of this
section).  $\mathcal {R}$ defines the distance from the origin of
the coordinate system to the point on the nuclear surface
\begin{eqnarray}
\mathcal {R} (\theta)=c_0    R \, [1+  \beta_{2} Y_{20}(\theta) +
\beta_{4} Y_{40}(\theta) +...],
\end{eqnarray}
with the scale factor $c_0$ which represents the effect of
incompressibility of nuclear matter in the nucleus and is
determined by the so-called constant volume condition
\cite{Cwoik}. $Y_{lm}(\theta, \phi)$ are the spherical harmonics.
$R=r_0 A^{1/3}$ and $a$ denote  the radius and surface diffuseness
of the single particle potential, respectively. Here, we assume
and set the radius and diffuseness of the single particle
potential of protons equal to those of neutrons for simplicity.
For protons the Coulomb potential is additionally involved (see
\cite{Cwoik} for details).

\begin{center}
\textbf{D. Symmetry Potential and Symmetry Energy Coefficient}
\end{center}

The relation between the isospin-asymmetric part $V_s$ of the
single particle potential depth in the microscopic part and the
symmetry energy coefficient in the macroscopic part is
investigated based on the Skyrme energy density functional
together with the ETF approach. In this approach, the central
one-body potential is described by $V_q=\frac{\delta \varepsilon
({\bf r})}{\delta \rho_q ({\bf r}) }$ with the energy density
functional $\varepsilon ({\bf r})$ (see Eq.(9) in Ref.
\cite{Bart02} for details).  The difference between the neutron
($q=n$) and proton ($q=p$) potentials of nuclear matter is written
as
\begin{eqnarray}
V_n-V_p &=& 2 B_2 \rho \delta + 2 B_8 \rho^{\alpha+1} \delta + B_4
(\tau_n-\tau_p) \nonumber \\
 & = & 2 B_2 \rho \delta + 2 B_8 \rho^{\alpha+1} \delta + B_4
c_k \rho^{5/3} \delta + \mathcal {O} (\delta^3)
\end{eqnarray}
with the kinetic energy density $\tau_q$ which can be expressed as
$\tau_q=\frac{3}{5} (3\pi^2)^{2/3}\rho_q^{5/3}$ in the
Thomas-Fermi approximation, the isospin asymmetry
$\delta=(\rho_n-\rho_p)/\rho$, and the coefficient $c_k=\left (
 3 \pi^2/2  \right )^{2/3}$. $B_2$, $B_8$ and $B_4$ (notations
in \cite{Bart02}) are some combinations of Skyrme parameters,
given by $B_2=-\frac{1}{2}t_0(\frac{1}{2}+x_0)$,
$B_8=-\frac{1}{12} t_3 (\frac{1}{2}+x_3)$ and $B_4=-\frac{1}{4} [
t_1 (\frac{1}{2}+x_1) - t_2 (\frac{1}{2}+x_2) ] $. The symmetry
potential $V_{sym}$ may be written as
\begin{eqnarray}
V_{sym} = \frac{V_n-V_p}{2 \delta} = B_2 \rho  +  B_8
\rho^{\alpha+1}  +   \frac{1}{2}B_4 c_k \rho^{5/3}  + \mathcal {O}
(\delta^2).
\end{eqnarray}
\begin{figure}
\includegraphics[angle=-0,width= 1\textwidth]{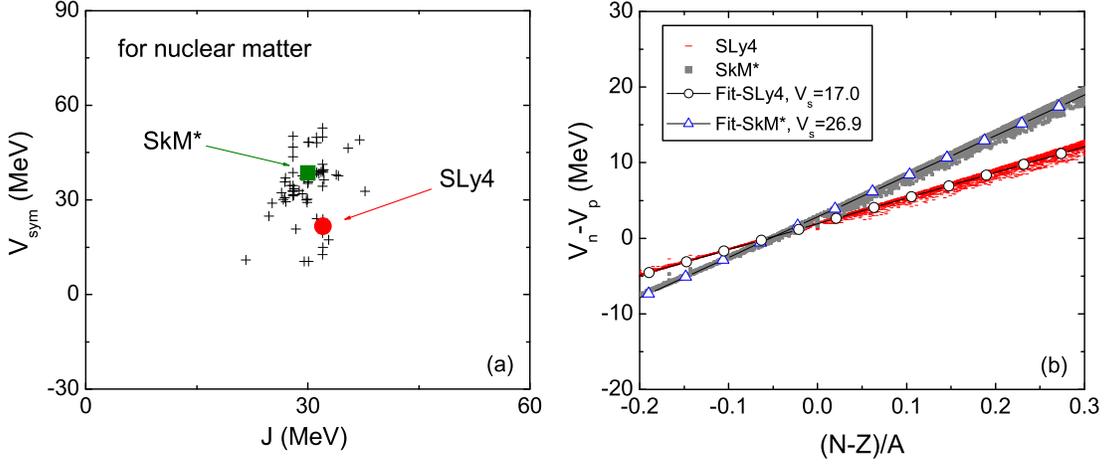}
 \caption{(Color online) (a) Symmetry potential $V_{sym}$ of nuclear matter with 78 Skyrme
 forces.  (b) The difference between the potential depth of neutrons and of protons as a function of mass asymmetry.
The circled curve and the triangled curve denote the results
fitted to the calculated results with SLy4 (short red dashes) and
SkM* force (small gray squares) for a large number of nuclei,
respectively.}
\end{figure}
The symmetry energy coefficient of nuclear matter $J$ is written
as \cite{Chab1}
\begin{eqnarray}
J= \frac{1}{2} B_2 \rho + \frac{1}{2} B_8 \rho^{\alpha+1} -
\frac{1}{24} \Theta_s c_k \rho^{5/3} + \frac{1}{3} \left (
\frac{\hbar^2}{2M}  \right ) c_k \rho^{2/3}
\end{eqnarray}
with $\Theta_s=3 t_1 x_1 - t_2 (4+5 x_2)$. The $\Theta_s$ term and
the last term of Eq.(16) give the contributions of the
effective-mass \cite{Bart02} and the kinetic energy to the $J$,
respectively. From the above equations for $V_{sym}$ and $J$, one
can get the relation between them,
\begin{eqnarray}
J= \frac{1}{2} V_{sym}  - \frac{1}{24} [\Theta_s+6 B_4]   c_k
 \rho^{5/3} + \frac{1}{3} \left ( \frac{\hbar^2}{2M}
\right ) c_k \rho^{2/3}.
\end{eqnarray}
A similar equation is previously proposed in \cite{Bar77} based on
perturbation theory,
\begin{eqnarray}
J=  \frac{1}{2} V_{sym}(k_{\rm F}) + \frac{1}{6} k_{\rm F} \left [
\frac{\partial V_0(k_m)}{\partial k_m} \right ]_{k_m=k_{\rm F}} +
\frac{1}{3}  \left ( \frac{\hbar^2 }{2 M} \right ) k_{\rm F}^2.
\end{eqnarray}
Due to the uncertainty of choosing the interaction parameters,
there exists a large uncertainty for the value of $V_{sym}$ in
different models. In Fig.3(a), we show the calculated symmetry
potential $V_{sym}$ of nuclear matter with 78 Skyrme forces. The
$V_{sym}$ has a value of about $10 \sim 50 $ MeV according to the
calculations. Brueckner-Hartree-Fock calculations show that the
value of $V_{sym}$ is about 25 MeV \cite{Zuo}. These calculations
indicate that the value of $V_{sym}$ is comparable to that of the
symmetry energy coefficient $J$ which is about 30 MeV.

For finite nucleus, the isospin-asymmetric part $V_s$ of the
single particle potential should be slightly different from the
value of $V_{sym}$. With the density distributions of nuclei
obtained in \cite{Liu06}, we calculate the potential depth of
protons and neutrons for a large number of nuclei. We find that
the difference $V_n-V_p$ increases linearly with the isospin
asymmetry $I$ (see Fig.3(b)). The average value for the
isospin-asymmetric part $V_s$ can be obtained by linearly fitting
the calculated results. The obtained values of $V_s$ are 17.0 and
26.9 MeV with SLy4 \cite{Chab} and SkM* \cite{Bart82} force,
respectively. In \cite{Jeuk}, the authors found that the
experimental Fermi energies of a number of magic nuclei can be
well described with a value of 23.2 MeV for $V_s$. The $a_{sym}$
in this work has a value of about $23\sim 24$ MeV for heavy nuclei
which is roughly comparable to the obtained values of $V_s$. In
the first round of searching for the optimal parameters of the
proposed mass formula, we treat $V_s$ as a free parameter and find
that the obtained value of $V_s$ is very close to that of
$a_{sym}$. So we empirically set and assume $V_s\approx a_{sym}$
in the improved mass formula for simplification.

\begin{center}
\textbf{E. Model Parameters}
\end{center}

\begin{table}
\caption{ Model parameters of the mass formula. }
\begin{tabular}{cc }
\hline\hline
  parameter                        & ~~~ WS~~~ \\ \hline
 $a_v  \; $ (MeV)                  &  $-15.5841$      \\
 $a_s \; $  (MeV)                  &   18.2359      \\
 $a_c \; $ (MeV)                   &   0.7173      \\
 $c_{sym} $(MeV)                   &   29.2876        \\
 $\kappa \;  $                     &   1.4492      \\
 $a_{  pair} $(MeV)                &   $-5.5108$       \\
 $g_1 $                            &   0.00862       \\
 $g_2 $                            &  $-0.4730$        \\
 $c_1  \; $                        &  0.7274        \\
 $V_0$ (MeV)                       &  $-47.4784$      \\
 $r_0$ (fm)                        &  1.3840           \\
 $a $ (fm)                       &  0.7842        \\
 $\lambda_0$                       &  26.3163     \\

 \hline\hline
\end{tabular}
\end{table}

From the above discussions, one can see that the macroscopic and
microscopic parts in the proposed mass formula are closely
connected to each other through the coefficient $a_{sym}$ of the
symmetry energy and  other isospin dependent model parameters. The
number of model parameters is considerably reduced  compared with
the finite range droplet model (FRDM) in which the number of
parameters is about 31 \cite{Lun03}. Here, we have 13 independent
parameters $a_v $, $a_s $, $a_c $, $c_{sym}$, $\kappa$,
$a_{pair}$, $g_1$, $g_2$, $c_1$, $V_0$, $r_0$, $a $, $\lambda_0$
for the nuclear mass. By varying these parameters and searching
for the minimal deviation of the 2149 nuclear masses from the
experimental data, we obtain a parameter set labeled as WS which
is listed in Table 1. To find the minimal energy $E (A,Z,\beta)$
with respect to a set of deformation parameters for a given
nucleus, the downhill searching method is adopted. We re-execute
the downhill algorithm for several times starting from different
initial deformation parameters in order to find the lowest energy
of a nucleus from some possible local minima on the energy surface
$E (A,Z,\beta)$. In the calculation of nuclear masses with the
obtained binding energies, the electron binding energies are not
included. In the parameter searching procedure, the downhill
searching method and the simulated annealing algorithm \cite{SA}
are incorporated. The former is used for the parameters of the
microscopic part, while the latter is for the macroscopic part.

\begin{center}
\textbf{III. RESULTS AND DISCUSSION}
\end{center}

In this section, we first show the calculated rms deviations of
the nuclear masses and of the neutron separation energies. In
addition, the change of magic number in light neutron-rich nuclei
and the shape coexistence phenomena for some nuclei have been
checked with the model. Then, the shell corrections of super-heavy
nuclei and the location of the center area of the super-heavy
island are investigated with the proposed mass formula.

\bigskip

\begin{center}
\textbf{A. Test of the Model}
\end{center}

\begin{table}
\caption{ rms $\sigma$ deviations between data AME2003 \cite{Audi}
and predictions of several models (in MeV). The line $\sigma (M)$
refers to all the 2149 measured masses, the line $\sigma (S_n)$ to
the 1988 measured neutron separation energies $S_n$. The
calculated masses with FRDM are taken from \cite{Moll95}. The
masses with HFB-14 and HFB-17 are taken from \cite{HFB14} and
\cite{HFB17}, respectively.}
\begin{tabular}{ccccc}
 \hline\hline
     & ~FRDM~ & HFB-14 & HFB-17 & ~~WS~~ \\
\hline
 $\sigma  (M)$      & $0.656$ & $0.729$ & $0.581$  & $0.516  $ \\
 $\sigma  (S_n)$    & $0.399$ & $0.598$ & $0.506$  & $0.346  $\\
  $N_p$             &   31    &   24      & 24      & 15 \\

 \hline\hline
\end{tabular}
\end{table}

The corresponding rms deviations of nuclear masses for the 2149
measured nuclei with the parameter set WS is listed in Table 2. In
addition, the results of FRDM and Hartree-Fock Bogoliubov (HFB-14
\cite{HFB14} and HFB-17 \cite{HFB17}) are also listed for
comparison. $N_p$ denotes the corresponding number of parameters
used in each model. Compared with the FRDM, the rms error for the
2149 nuclear masses is considerably reduced  with WS, from 0.656
to 0.516 MeV. The number of parameters in the model is reduced
from 31 to 15 (including the two parameters $\gamma$ and $p$ used
in the Strutinsky procedure). One should note that several (about
12) of the FRDM parameters were prefixed by considerations other
than mass-like data before making the fit \cite{Moll95,Lun03} and
also that the fit in the FRDM included data on fission barriers in
addition to masses. In this work, only the precisely measured
nuclear masses are involved in the fit. Compared with the standard
Hartree-Fock Bogoliubov (HFB) approach, the CPU time used in the
calculation of nuclear mass table is much shorter with the
proposed mass formula. The obtained rms error for the 1988
measured neutron separation energies $S_n$ with our model is
obviously smaller than those of HFB calculations
\cite{HFB14,HFB17}.

\begin{figure}
\includegraphics[angle=-0,width=0.7\textwidth]{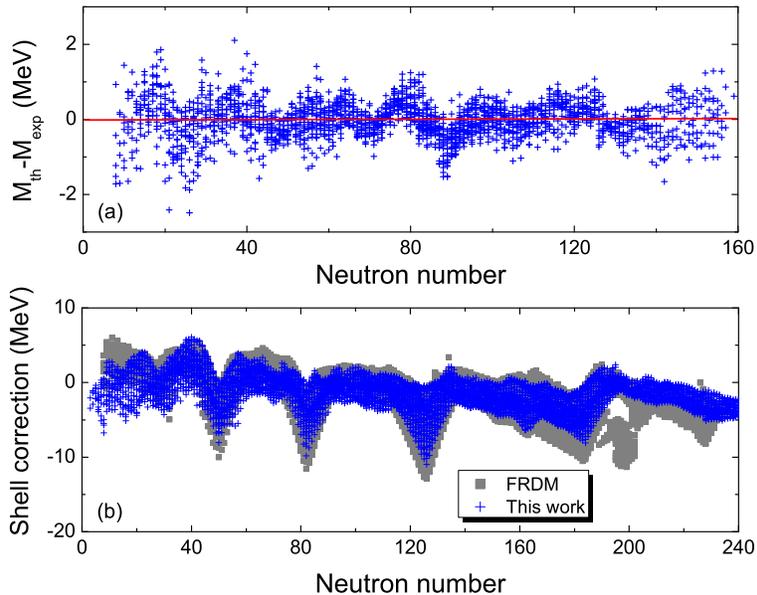}
 \caption{(Color online) (a) Deviations between the calculated
 nuclear masses from the experimental data.
 (b) Calculated shell corrections $\Delta E$ of nuclei (crosses). The squares denote the microscopic energy of nuclei with the FRDM
model (column $E_{mic}$ of the table of Ref.\cite{Moll95}). }
\end{figure}

Fig.4(a) shows the deviations between the calculated  nuclear
masses in this work from the experimental data. In Fig.4(b), we
show the calculated shell corrections $\Delta E$ of nuclei with
our model and the microscopic energy (mainly including the shell
correction and the deformation energy) obtained in the
finite-range droplet model. For intermediate and known heavy
nuclei, the results of the two approaches are comparable and both
of them reproduce the known magic numbers very well. The
deviations are large for light nuclei and super-heavy nuclei. Our
calculations show that the shell corrections of nuclei with about
$N=16$ are much larger (in absolute value) than those from the
FRDM. As an example, the shell correction of $^{24}$O is
calculated and has a value of $-4.6$ MeV with WS. Experimentally,
it is thought that $^{24}$O is a doubly magic nucleus from the
observed decay energy spectrum and the high-lying first excited
2$^+$ state (above 4.7 MeV) \cite{Hoff}, which is consistent with
our calculations. The obtained shell corrections with WS for
$^{20}$C, $^{22}$C \cite{Tan10} and $^{23}$N are $-4.2$, $-5.1$
and $-4.3$ MeV, respectively. Some theoretical and empirical
studies \cite{Gupta, Hoff} have shown that in the neutron-rich
nuclei the magic numbers such as $N=14$ or 16 can arise, which is
in agreement with our calculations. It is known that the shell
correction strongly depends on the single particle potential
adopted. The isotopic dependence of the spin-orbit strength and
the symmetry potential adopted in this work is different from that
in the FRDM, which leads to the different shell correction from
the two models. Our results for the neutron-rich nuclei with about
$N=16$ look more reasonable qualitatively.

\begin{figure}
\includegraphics[angle=-0,width=0.75\textwidth]{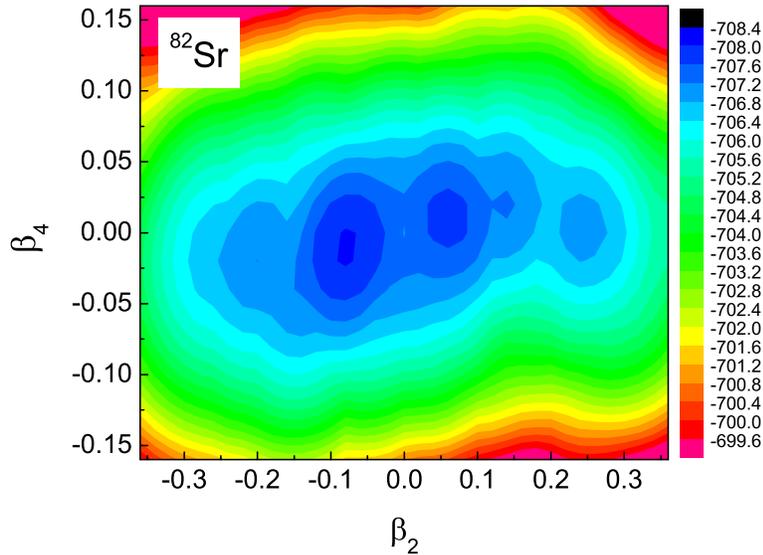}
 \caption{(Color online) Potential energy surface $E
(\beta_2, \beta_4)$ of $^{82}$Sr.}
\end{figure}

To further test the model, we study the potential energy surface
$E (\beta_2, \beta_4)$ of some nuclei. In \cite{Petr}, the authors
observed the shape coexistence phenomena for nuclei $^{82}$Sr and
Kr isotopes from the low and high spin states. The shape
coexistence phenomena of these nuclei could be observed from the
corresponding potential energy surface. In Fig.5, we show the
calculated potential energy surface of $^{82}$Sr. The coexistence
of oblate and prolate deformed configurations can be clearly
observed. The similar coexistence phenomena for Kr isotopes can
also be observed with our model.

\begin{center}
\textbf{B. Shell Corrections of Super-heavy Nuclei}
\end{center}

\begin{table}
\caption{ Shell corrections of some nuclei (in MeV). The data of
FRDM are taken from the microscopic energies $E_{mic}$ of the
table of Ref. \cite{Moll95}.}
\begin{tabular}{ccccccccccccc}
 \hline\hline

   &$^{16}$O & $^{24}$O & $^{40}$Ca & $^{48}$Ca  & $^{90}$Zr & $^{132}$Sn & $^{208}$Pb & $^{270}$Hs & $^{288}$114 &$^{292}$114 & $^{298}$114 & $^{294}$116\\
\hline
 WS  & $-0.7$ & $-4.6$ & $2.0$ & $-1.2$ & $-1.3$& $-9.8$  & $ -11.0 $ & $-6.4$ & $-5.3$ &$-6.1$ & $-6.1$  & $-6.2$  \\
 FRDM & $2.1$ & $0.3$  & $2.3$ & $ 0.1$ & $-1.6$& $-11.6$ & $ -12.8 $ & $-6.5$ & $-7.8$ &$-8.9$ & $-7.6$  & $-8.7$ \\

 \hline\hline
\end{tabular}
\end{table}

The precise calculation for the shell corrections of super-heavy
nuclei is of great importance for the synthesis of new super-heavy
nuclei, especially for the prediction of the location of the
super-heavy island. Furthermore, the fission barriers of
super-heavy nuclei are roughly estimated by the values of the
corresponding shell corrections \cite{Shen08,Mull86} since the
macroscopic fission barriers disappear at super-heavy region in
general. It is known that the fission barrier is a very sensitive
parameter in the realistic calculations for the survival
probabilities of the produced compound nuclei. It is therefore
necessary to investigate the shell corrections of super-heavy
nuclei.

\begin{figure}
\includegraphics[angle=-0,width=1.0\textwidth]{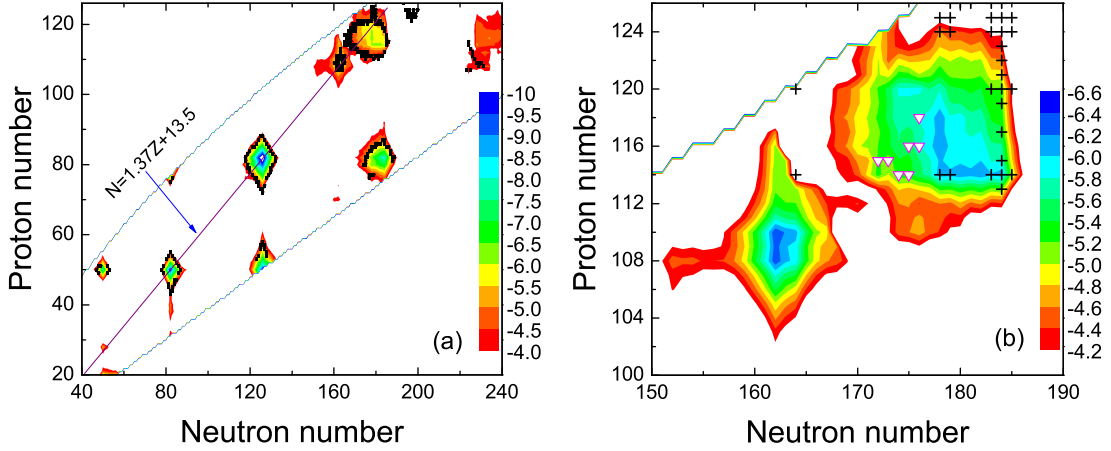}
 \caption{(Color online) (a) Shell correction energies $\Delta E$ of nuclei.
 The black squares denote the nuclei with microscopic energies of $-(6\sim7)$ MeV in the FRDM calculations.
  The straight line passes through the areas with the known heavy magic
  nuclei. (b) Shell correction energies of nuclei in super-heavy
  region. The crosses denote the nearly spherical nuclei (calculated $|\beta_2| \le 0.01$) and the triangles denote the synthesized super-heavy nuclei in the "hot" fusion reactions
 \cite{Ogan04,Ogan06}.
}
\end{figure}

In Table 3, we list the calculated shell corrections $\Delta E$ of
some nuclei. The corresponding microscopic energy obtained in the
FRDM are also listed for comparison. For super-heavy nuclei such
as nucleus $^{292}$114, the microscopic energies obtained with the
FRDM are much lower (absolute value larger about $2\sim3$ MeV)
than our calculated $\Delta E$. Because these nuclei are (nearly)
spherical in shape according to the calculations. It follows that
the deviations of shell energies between the two models are about
$2\sim3$ MeV for nuclei around $^{292}$114. Because the shell
correction can not be measured directly, it is still difficult to
quantitatively compare the reliability of model through the
calculated shell corrections of nuclei. In addition, we study the
central area of the super-heavy island based on the calculated
shell energies. Fig.6 shows the contour plot of the calculated
shell correction energies of heavy nuclei. The black squares in
sub-figure (a) denote the nuclei with microscopic energies of
$-(6\sim7)$ MeV in the FRDM calculations.  One can see that both
of models give similar magic numbers for heavy nuclei. Fig.6.(b)
shows the shell corrections of nuclei in the super-heavy region.
The crosses denote the calculated nearly spherical nuclei
($|\beta_2|\le 0.01$) with WS. The predicted super-heavy island
according to the obtained shell corrections of nuclei looks flat.
Along nuclei with $Z=114$ or $N=178$, one can see a slightly
deeper valley in the contour plot of shell corrections. The
calculated deformations of nuclei demonstrate that the nuclei with
$N=184$ are (nearly) spherical in shape. However, the maximum
shell correction occurs at $N=178$ instead of $N=184$, which is
consistent with the results in \cite{Fiset,Mull86}. The analysis
about the shift of the shell correction from $N=184$ to $N=178$ is
given in Ref. \cite{Mull86}. According to the calculations, the
super-heavy nuclei $^{288, 289}$114 produced in the "hot" fusion
reaction $^{48}$Ca+$^{244}$Pu \cite{Ogan04} (the corresponding
compound nucleus is $^{292}$114) are close to this central area of
the island. The half-lives of these nuclei are in the order of
seconds \cite{Ogan04}, which is much shorter than those of known
stable nuclei. The measured short half-lives of nuclei in
super-heavy region seem to indicate that the shell corrections of
these nuclei are probably not very large.
\begin{figure}
\includegraphics[angle=-0,width=1.0\textwidth]{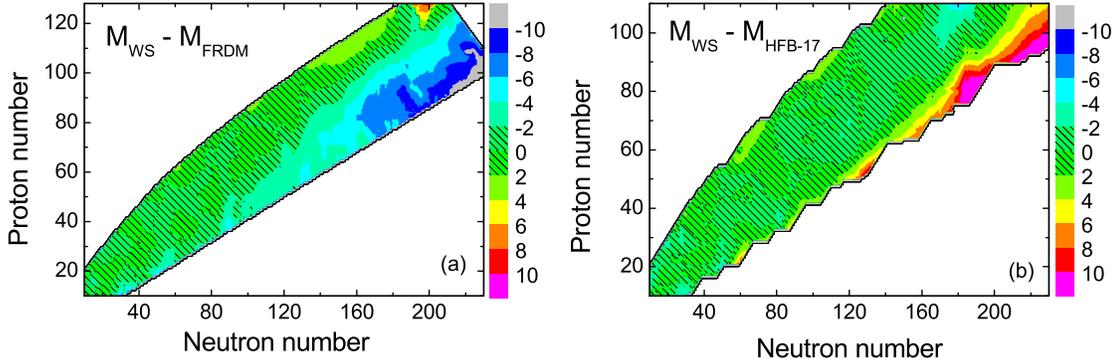}
 \caption{(Color online) Deviations of the calculated nuclear masses in this work from the results of FRDM (a) and HFB-17 (b), respectively. The
calculated masses with FRDM and HFB-17 are taken from
\cite{Moll95} and \cite{HFB17}, respectively. The shades denote
the region with deviations smaller than 2 MeV.}
\end{figure}

Fig.7 shows the deviations of the calculated nuclear masses with
the proposed model from the results of FRDM and HFB-17. The shades
denote the region with deviations smaller than 2 MeV. The results
for highly neutron-rich heavy nuclei from the three models have
large deviations. The results of our model are relatively close to
those of HFB-17 for most nuclei.

\begin{center}
\textbf{IV. SUMMARY}
\end{center}

In this paper we proposed a semi-empirical nuclear mass formula
based on the macroscopic-microscopic approach. The isospin effects
in both macroscopic and microscopic part of the formula are
self-consistently considered, with which the number of model
parameters is considerably reduced compared with the finite range
droplet model and the rms deviation of the calculated masses from
the 2149 measured nuclear masses is reduced by $21\%$ and falls to
0.516 MeV. The CPU time used in the calculation of the nuclear
masses for the whole nuclear chart is much shorter than that with
the microscopic mass formula models. At the same time the
consistency of the model parameters between the macroscopic and
microscopic parts greatly promotes the credibility of
extrapolations in the macroscopic-microscopic approach.

In order to extend the mass formula to super-heavy nuclei and the
nuclei far from the $\beta$-stability line, we pay a special
attention to study the isospin and mass dependence of the model
parameters including symmetry energy coefficient and the symmetry
potential, etc. Those studies are based on the Skyrme energy
density functional approach together with the extended
Thomas-Fermi approximation. Since more sufficiently considering
the isospin effects of the model parameters the formula could
systematically study super-heavy nuclei and the nuclei far from
the $\beta$-stability line.

To further test the model, the appearance of new magic number
$N=16$ in light neutron-rich nuclei and the shape coexistence
phenomena for some nuclei have been examined with the model. Our
results are in good agreement with some experimental and
theoretical studies. The predicted super-heavy island according to
the obtained shell corrections of nuclei looks flat. Along nuclei
with $Z=114$ or $N=178$, we find a relatively deeper valley in the
contour plot of shell corrections. The shell corrections of nuclei
around $^{292}$114 are about $-6$ MeV and much smaller (in
absolute value) than the corresponding results from the
finite-range droplet model. The calculated nuclear masses for
highly neutron-rich heavy nuclei from the three different models
have large deviations. The results of our model are relatively
close to those of HFB-17 for most nuclei.

\begin{center}
\textbf{ACKNOWLEDGEMENTS}
\end{center}

We thank Prof. W. Scheid  for a careful reading of the manuscript.
We also thank Prof. Zhuxia Li and an anonymous referee for
valuable suggestions. This work was supported by National Natural
Science Foundation of China, Nos 10875031, 10847004 and 10865002.
The obtained mass tables with the proposed formula are available
from http://www.imqmd.com/wangning/publication.html


\begin{thebibliography}{99}

\bibitem{Lun03} D. Lunney, J.M. Pearson, C. Thibault, Rev. Mod. Phys. \textbf{75} (2003)
1021.

\bibitem{Moll95} P. M\"oller, J. R. Nix, et al., At. Data and
Nucl. Data Tables \textbf{59} (1995) 185.

\bibitem{HFB14} S. Goriely, M. Samyn and J. M. Pearson, Phys. Rev. C \textbf{75} (2007) 064312.
\bibitem{HFB17} S. Goriely, N. Chamel and J. M. Pearson, Phys. Rev. Lett. \textbf{102} (2009)
152503.

\bibitem{DZ} J. Duflo and A. P. Zuker, Phys. Rev. C \textbf{52} (1995) 23.


\bibitem{Audi} G. Audi, A.H. Wapstra and C. Thibault, Nucl. Phys. A \textbf{729} (2003)
337.
\bibitem{Bart02} J. Bartel and K. Bencheikh, Eur. Phys. J, A\textbf{14} (2002 179).

\bibitem{Liu06} Min Liu, Ning Wang, et al., Nucl.
Phys. A \textbf{768} (2006) 80.
\bibitem{Cohen} S. Cohen, F. Plasil and W. Swiatecki, Ann. Phys. \textbf{82} (1974)
557.

\bibitem{Heyde} K. Heyde, \emph{Basic Ideas and Concepts in
Nuclear Physics} (IOP, Bristol, 1999).


\bibitem{Tem} J. Mendoza-Temis, J. G. Hirsch, A. P. Zuker, arXiv:0912.0882.

\bibitem{Pear01} J. M. Pearson, Hyperfine Interactions \textbf{132} (2001)
59.

\bibitem{Kirson} Michael W. Kirson, Nucl. Phys. A \textbf{798} (2008) 29.


\bibitem{Sam} S. K. Samaddar, J. N. De, et al., Phys. Rev. C
\textbf{76} (2007) 041602.

\bibitem{Bart82} J. Bartel, Ph. Quentin, et al., Nucl. Phys. A  \textbf{386} (1982)
79.


\bibitem{Strut} V. M. Strutinsky  and F. A. Ivanjuk, Nucl. Phys. A \textbf{255} (1975)
405.

\bibitem{Sag09} R. N. Sagaidak and A. N. Andreyev, Phys. Rew. C \textbf{79} (2009)
054613.


\bibitem{Shen08} C. Shen, Y. Abe, et al., Int. J. Mod. Phys. E \textbf{17} (2008) 66.

\bibitem{Cwoik} S. Cwoik, J. Dudek, et al., Comp. Phys. Comm. \textbf{46} (1987) 379.

\bibitem{Chab1} E. Chabanat, P. Bonche, et al., Nucl. Phys. A \textbf{627} (1997)710.


\bibitem{Bar77}  K. M. Khanna, D. Jairath and P. K. Barhai, Czech. J. Phys. B \textbf{27} (1977) 498.

\bibitem{Zuo} W. Zuo, L. G. Cao, B. A. Li, et al., Phys. Rev. C \textbf{72}
(2005) 014005.

\bibitem{Chab} E. Chabanat, P. Bonche, et al, Nucl. Phys. A \textbf{635}
(1998) 231.


\bibitem{Jeuk} J. P. Jeukenne, C. Mahaux, and R. Sartor, Phys. Rev. C \textbf{43}
(1991) 2211.

\bibitem{SA} A. Corana, M. Marchesi, et al., ACM Transactions on Mathematical
Software, \textbf{13}  (1987) 262.

\bibitem{Hoff} C. R. Hoffman, T. Baumann, D. Bazin, et al., Phys. Lett. B
\textbf{672} (2009) 17.

\bibitem{Tan10} K. Tanaka, T. Yamaguchi, et al., Phys. Rev. Lett. \textbf{104}
(2010) 062701.

\bibitem{Gupta} Raj K Gupta, et al., J. Phys. G: Nucl. Part.
Phys. \textbf{32} (2006) 565, and references therein.

\bibitem{Petr} A. Petrovici, et al., Prog. Part. Nucl. Phys. \textbf{43} (1999)
485.

\bibitem{Mull86} P. M\"oller, G. A. Leander and J. R. Nix, Z. Phys. A
\textbf{323} (1986) 41.

\bibitem{Fiset} E. O. Fiset, and J. R. Nix, Nucl. Phys. A \textbf{193} (1972)
647.



\bibitem{Ogan04} Yu. Ts. Oganessian, V. K. Utyonkov, et al., Phys. Rev. C \textbf{69}
 (2004) 054607.

\bibitem{Ogan06} Yu. Ts. Oganessian, V. K. Utyonkov, et al., Phys. Rev. C \textbf{74}
 (2006) 044602.





\end{thebibliography}
\end{document}